# Unraveling Energetic Disorder in Organic Bulk Heterojunction Photovoltaics by Capacitance-Voltage Spectroscopy

Xixiang Zhu[1], Kai Wang[1]*, Changfeng Han[1], Qin Yang[1], Xiaojuan Sun[1], Haomiao Yu[1], Ming Shao[1], Fujun Zhang[1] and Bin Hu[1,2]*

1. Key Laboratory of Luminescence and Optical Information, Ministry of Education, Beijing Jiaotong University, Beijing 100044, China
2. Department of Materials Science and Engineering, University of Tennessee, Knoxville, TN 37996, USA

Correspondence should be addressed to:
Kai Wang
E-Mail: kaiwang@bjtu.edu.cn
Bin Hu
E-Mail: bhu@utk.edu

**ABSTRACT**

Organic semiconductors possess an intrinsic energetic disorder characteristic, which holds an exceptionally important role for understanding organic photovoltaic (OPV) operation and future optimization. We performed illumination intensity dependence of capacitance-voltage (*C-V*) measurements in PIDTDTQx:PC$_{70}$BM based organic bulk heterojunction (BHJ) photovoltaics in working conditions. Energetic disorder profiles for the photo-active layer, PIDTDTQx:PC$_{70}$BM, changed significantly when different interfaces were involved. The effects of energetic disorder that could be reflected from *C-V* profiles are incorporated through an exponential or Gaussian model of density of states (*DOS*), or a combination of these two. Results underlie that an identical organic blend in BHJ solar cells exhibits different energetic disorder when it interacts with various interfaces. It may, thus, has a certain impact on OPV performances, such as open-circuit voltage (*V$_{oc}$*). Our study provides device physicists a different perspective view for tailoring the organic energetic disorder parameter via interfaces in order to enhance photo-electron conversion efficiencies (*PCE*).

**INTRODUCTION**

Organic bulk heterojunction (BHJ) photovoltaics which are formed by an interpenetrating blend of photon-actively conjugated polymers and fullerene derivatives (or non-fullerene derivatives) have emerged as a promising candidate for organic photovoltaics (OPVs).[1-5] It has witnessed significant improvements of power conversion efficiencies (PCE) with the highest record of more than 10% at present.[5-10] One approach to obtain desirable PCE is to increase the open circuit voltage ($V_{oc}$). It is rarely to achieve $V_{oc}$ of more than 1 V even in many high performance organic BHJ solar cells, since $V_{oc}$ is determined by a few factors including donor-acceptor energy gaps, illumination intensities, charge-carrier recombination rates, contact work functions and energetic disorder.[11-16] Among these, one of the most decisive parameters that has been less discreetly considered by device physicists is the amount of energetic disorder due to energy level broadening effect in organic solid films.

The energetic disorder of an organic BHJ blend that depicts highest occupied molecular orbital (HOMO) of a donor and lowest unoccupied molecular orbital (LUMO) of an acceptor does not have a well-defined energetic onset.[17,18] S-Figure 1(a) of the supplementary information schematically shows that the tails of energy states for both LUMO and HOMO extend into the energy gap. Theoretical and experimental studies have revealed that such effect is remarkable and its strength varies differently in various organic semiconductors and film preparations.[19-21] Smart ways to describe energetic disorder profiles and shapes of density of states (*DOS*) are based on exponential (S-Figure 1(b)) or Gaussian (S-Figure 1(c)) distributions – or a combination of these two.[11,22-24] Full pictures of *DOS* for describing BHJ solar cells under working conditions are crucial to understand reasons of $V_{oc}$ losses. Nevertheless, it remains challenging to experimentally directly image *DOS*. In order to facilitate a narrow focus on the energetic disorder in organic BHJ solar cells, PIDTDTQx:PC$_{70}$BM based organic photovoltaics together with some closely relevant device configurations, such as without electron transport layer (i.e., PFN) or hole transport layer (i.e., PEDOT:PSS) or even without both of them, were systematically investigated using the non-destructive capacitance-voltage (*C-V*) spectroscopic technique. Some experimental details can be found in the supporting information (S-Figure 2). In this work, we would like to understand, to which extend, the energetic disorder of the photo-active layer (i.e., PIDTDTQx:PC$_{70}$BM) would be influenced for organic BHJ photovoltaics under working conditions. We have experimentally found that the energetic disorder can be elucidated from *C-V* characteristics. Some remarkable changes have been observed when the photo-active layer was involved with different interfaces.

**RESULTS and DISCUSSION**

S-Figure 2 of the supplementary information shows that the basic *J-V* measurements for four different BHJ devices under illuminations. They are ITO/PEDOT:PSS/PIDTDTQx:PC$_{70}$BM/PFN/Al (black), ITO/PEDOT:PSS/PIDTDTQx:PC$_{70}$BM/Al (red), ITO/PIDTDTQx:PC$_{70}$BM/PFN/Al (green) and ITO/PIDTDTQx:PC$_{70}$BM/Al (blue), respectively. The results for their photovoltaic performances have been summarized in S-Table 1 of the supplementary information. The active areas/effective areas are approximately 4 mm$^2$ for all the organic BHJ photovoltaic devices. The ITO/PEDOT:PSS/PIDTDTQx:PC$_{70}$BM/PFN/Al produces the greatest PCE of 6.24%; while, the $V_{oc}$ and $J_{sc}$ are 0.86 V and 13.71 mA/cm$^2$ respectively. When the PFN or/and PEDOT:PSS layers are removed, the corresponding $V_{oc}$, $J_{sc}$, *FF* and *PCE* decrease for the rest of the organic BHJ devices. We will demonstrate below that the absence of the interfacial layers, indeed, have a significant impact on the energetic disorder for the active layer PIDTDTQx:PC$_{70}$BM, it may play a crucial role for the reduction of $V_{oc}$.

Figure 1(a) shows the *C-V* characteristics which were measured at several different illumination intensities for the device comprising ITO(glass)/PEDOT:PSS/PIDTDTQx:PC$_{70}$BM/PFN/Al. All the measurements exhibit broad *C-V* bands at some positive bias voltage. The capacitance that varies at different sweeping bias voltage is primarily ascribed to three capacitive effects. The one that was measured in dark displays almost a constant value at relatively large negative bias voltage (i.e., electrons flow from occupied energy states of ITO to unoccupied energy states of Al). Such capacitive effect originates from the geometric capacitance ($C_{geo} = (A\varepsilon_r\varepsilon_0)/L$), in which, *A* is the effective area for charge transport, $\varepsilon_r$ is the relative static permittivity, $\varepsilon_0$ is the permittivity of free space, and $L$ is the separation between the top and bottom electrodes.

Despite this, the device contains two different interfaces that are ITO/PEDOT:PSS and Al/PFN respectively. It is known that an interfacial depletion region can be created at the metal-organic interface and it generates the so-called surface-depletion capacitance ($C_{sc}$). Without external stimuli, the metal-organic interface results in an alignment of their Fermi-energies in order to reach an energetic equilibrium at a fixed temperature. This, as a consequence, leads to an interfacial band bending.[25] At some moderately negative and low positive bias voltage (i.e., still far below $V_{oc}$), the applied bias voltage ($V_{app}$) can be used to modulate the width of the depletion zone by adding with the build-in potential ($V_{bi}$). The correlation of $C_{sc}$ with the difference between $V_{app}$ and $V_{bi}$ (also called flat-band potential $V_{fb}$) is $C_{sc} = (A\varepsilon_r\varepsilon_0)/w_0\sqrt{|V_{sc}|}$, in which, $V_{sc} = V_{app} - V_{bi}$, $w_0$ is the width of the surface depletion zone. Further increase of the bias voltage that exceeds $V_{bi}$ (but still much less than $V_{oc}$) continuously suppresses the depletion zone. Until $V_{bi}$ is completely suppressed, $V_{app}$ that approaches to $V_{oc}$ leads to the increase of the capacitance owing to charge accumulation at the organic-metal interface and continuous split of the quai-Fermi-energies levels in the donor-acceptor blend. The corresponding capacitance which is due to charge accumulation close to the quasi-Fermi-energies in the active layers is governed by the so-called chemical capacitance ($C_\mu$). Therefore, $C_\mu$ is proportional to the charge carrier density ($n$) with respect to the quasi-Fermi energies, and it can be expressed as,

$$C_\mu = Lq^2 \frac{dn}{dE_F} = Lq^2 g(E_F) \qquad (1)$$

in which, $q$ is the elementary charge, $E_F$ is Fermi-energy, and $g(E_F)$ represents *DOS* at Fermi-energy. The total charge carrier density can be expressed as an integration of the *DOS*,

$$n = \int g(E) f(E - E_F) dE \qquad (2)$$

where $f(E - E_F)$ represents the Fermi-Dirac distribution function. $C_\mu$ can then be further modified as,

$$C_\mu = \frac{q^2}{k_B T} \int g(E) f(E - E_F) [1 - f(E - E_F)] dE \qquad (3)$$

Thus, some changes of $g(E)$ can influence $C_\mu$. With this concept, two analytical models, (I) exponential distribution and (II) Gaussian distribution, can be used to analyze the *DOS* broadening effect and its profile. Below are the mathematical expressions for the corresponding modeled *DOS*,

$$g_{h/e}(E) = \frac{N_{t,h/e}}{E_t} e^{[\pm \frac{E - E_{HOMO,D/LUMO,A}}{E_t}]} \qquad (4)$$

$$g_{h/e}(E) = \frac{N_{h/e}}{\sigma \sqrt{2\pi}} e^{[-\frac{(\pm E \mp E_{HOMO,D/LUMO,A})^2}{2\sigma^2}]} \qquad (5)$$

in which, $N_{h/e}$ is the total hole or electron densities in the donor-acceptor blend respectively, $E$ denotes energy, $E_{HOMO,D}$ and $E_{LUMO,A}$ represent the donor HOMO energy level and acceptor LUMO energy level respectively, $E_t$ denotes the characteristic energy for the exponential tail distribution, and $\sigma$ quantifies the energetic disorder in the Gaussian distribution.

Figure 1(b) and (c) which show the two selected *C-V* curves corresponding to the dark and complete light exposure (i.e., 100 mW/cm$^2$ illumination) respectively for the device consists of ITO/PIDTDTQx:PC$_{70}$BM/PFN/Al. They were well fitted by equation 4 using the exponential function of *DOS*. Since the insulating PFN film was made very thin in this case (i.e., less than 10 nm), and the electronic charge transport is primarily dominated by multi-step electronic tunneling process, the conventional Mott-Schottky barrier cannot be formed at the PFN/Al interface. Therefore, $C_{sc}$ can be ignored, and the capacitive signal at positive bias is mainly due to the split of quasi-Fermi energies. Figure 1(b) shows the gradually increase of the capacitance between 0 V and 0.78 V. The same phenomenon was observed when the device was fully light exposed. $C_\mu$ arises due to accumulation of photo-generated charge carriers close to the quasi-Fermi-energy levels (Figure 1(c)). Upon photo-excitations, the total electron density was as high as ~10$^{17}$ cm$^{-3}$. The *C-V* measurements depict that the energetic profiles which are associated with *DOS* follow the exponential form exactly. A clear displacement of the *C-V* curves by approximately 0.20 V towards smaller positive bias voltage can be observed by comparing figure 1(b) and (c). Such effect has been previously reported due to charge occupancy of interfacial *DOS* upon photo-excitations.[26,27]

Figure 1(d) shows the same experiments for a modified device comprising ITO/PEDOT:PSS/PIDTDTQx:PC$_{70}$BM/Al, in which, the top metallic Al electrode is in direct contact with the thick photo-active layer PIDTDTQx:PC$_{70}$BM. The *C-V* bands which were measured at several different illumination intensities show a much less displacement by comparing with figure 1(a), since the different interface (i.e., PIDTDTQx:PC$_{70}$BM/Al) was involved. Two selected *C-V* bands of figure 1 (e) and (f) which correspond to the cases of without and with illumination reveal more steeper *C-V* profiles at some positive bias voltage close to $V_{oc}$ than those shown in figure 1(b) and (c). When the device was fully light exposed, a co-existence of the exponential and Gaussian profiles can be clearly observed from both experiment and curve fitting in figure 1(f). The Gaussian profile starts to play a role within the bias window between 0.6 V and 0.7 V. Such intriguing phenomenon has not been previously experimentally reported in organic BHJ solar cells by the *C-V* spectroscopy.

It may be also interesting to consider when the HTL such as PEDOT:PSS is removed. Figure 2(a) shows the corresponding *C-V* measurements at several different illumination intensities for a device consisting of ITO/PIDTDTQx:PC$_{70}$BM/PFN/Al. Surprisingly, all the *C-V* bands display the quasi-symmetrically and relatively narrow bands at large forward bias voltage by comparing with figure 1(a) and (d). The results may indicate that it is possible to tune the energetic disorder parameter for PIDTDTQx:PC$_{70}$BM when its solid thin film is prepared on a different material. As we can see from figure 2(b) and (c), the *C-V* characteristic which was measured under the illumination shows a broader band than the one was measured in the dark. The relative shift of the *C-V* bands is equal to 0.232 V. Similar *C-V* signals were detected for the device consists of ITO/PIDTDTQx:PC$_{70}$BM/Al, in which the photo-active layer PIDTDTQx:PC$_{70}$BM was fabricated in-between ITO and Al electrodes. The results of the light intensity dependence of the *C-V* bands were plotted in figure 2(d).

Clearly, with external light stimuli, the *C-V* profiles of figure 2(a) and (d) exhibit Gaussian distribution at large forward bias voltage close to $V_{oc}$. The *C-V* bands shift toward smaller positive bias voltage with the increase of the illumination intensities. However, the relative amount of displacement for the *C-V* peaks is different when the photo-active layers PIDTDTQx:PC$_{70}$BM is fabricated in different device configurations. In contrast, the Gaussian profile of figure 2(f) is broader than the one shown in figure 2(c), our fittings depict that it leads to an increase of disorder parameter σ from 205 meV to 220 meV.

In this article, the *C-V* profiles for the active layer, PIDTDTQx:PC$_{70}$BM, in the four device configurations are distinctly different. We proposed that they are attributed to the significant modifications of energetic disorder due to the energy broadening effect in the PIDTDTQx:PC$_{70}$BM organic blend. Such parameter is, in fact, an inherent property for organic semiconductors and their blends. In most circumstances, they may show some high degrees of energetic disorder compared to their highly ordered crystalline inorganic counterparts that are not significantly influenced by moderate variation of energetic disorder due to their higher charge carrier densities and crystallinity. The high charge carrier densities and crystallinity give rise to high recombination rates and some well-defined band-edges. It is evitable that the increase of energetic disorder elaborates the energy broadening effect and the falling of energy states within the band-gap. Figure 3 schematically illustrates the donor-acceptor energy states with the energy broadening effect at the vicinities of band-edges. Our *C-V* measurements are firmly correlated to *DOS* close to Fermi-energy at some large forward bias voltage; the profiles may exhibit an exponential or Gaussian distributions or a combination of these two. As we can see from figure 3, a continual increase of the forward bias voltage gives a rise to the continual splitting of quasi-Fermi energies, such as $E_F^n$ and $E_F^p$. As the energy splitting continuously happens from $E_F^p$ to $E_F^{p1}$ and $E_F^{p2}$, the quasi-Fermi energy may sweep along either exponential or Gaussian *DOS* or a combination of these two. The occupancy/accumulation of charge carriers at tail energy states can therefore be reflected from the *C-V* measurements.

As the results shown in figure 1 and 2, the exponential and Gaussian shaped *DOS* indicate the PIDTDTQx:PC$_{70}$BM based organic BHJ photovoltaics could have a high degree of energetic disorder depending on different adjacent layers. The exponential shaped *C-V* profile at positive bias voltage of figure 1(a) that corresponds to the largest PCE of 6.24% tells us the photo-electron generation process well overcomes the recombination process. After the PFN films is removed and when the device is under the light exposure, figure 1(d) shows that the *C-V* profiles change significantly judging from three aspects, (i) the shift of the *C-V* curves, (ii) the steep increase of the capacitance at positive bias voltage, and (iii) the supposition of both exponential and Gaussian shaped *DOS* at large positive bias voltage. The former two can be understood since PIDTDTQx:PC$_{70}$BM/Al interfacial *DOS* are modified, and the formation of $C_{sc}$ produced by $V_{sc}$. Here, we propose that the later one (i.e., iii) is closely attributed to a deep *DOS* that is the exponential distribution superposed on another shallow *DOS* that is the Gaussian distribution. Both of them have certain extensions within the bandgap forming tails of *DOS*. It may increase recombination rates due to trapped charge carrier to trapped charge carrier recombination process (i.e. process I of figure 3), and free charge carrier to trapped charge carrier recombination process (i.e. process II of figure 3); as a consequence, they cause a certain reduction of $V_{oc}$.[11,28,29] In spite of these two, we have also observed that the PEDOT:PSS, indeed, has remarkable impacts on the shape of the *DOS*. From figure 2 (a), the Gaussian shaped *DOS* starts to dominate for every *C-V* measurements for the device consists of ITO/PIDTDTQx:PC$_{70}$BM/PFN/Al. Similar phenomena have been observed for PIDTDTQx:PC$_{70}$BM sandwiched between ITO and Al. In both cases, the significant decrease of $V_{oc}$ may be accredited to different broadening strengths of *DOS*. Their corresponding FWHM of the Gaussian distributions elucidate the degree of energetic disorder; and in this case, figure 2(f) shows clearly a wider Gaussian distribution than the one shown in figure 2(c). A large degree of energetic disorder means a high density of traps. All the results indicate that the energetic disorder for PIDTDTQx:PC$_{70}$BM based organic BHJ solar cells could be much more sensitive to their adjacent layers. Therefore, to precisely understand and effectively manipulate the energetic disorder parameter for organic blends appear in organic BHJ solar cells may shed a new light on the PCE improvements.

In summary, we have demonstrated *C-V* measurements for organic BHJ photovoltaics consist of ITO(glass)/PEDOT:PSS/PIDTDTQx:PC$_{70}$BM/PFN/Al. The same measurements were performed after removing ETL (i.e., PFN), HTL (i.e., PEDOT:PSS) and both of them. From both experimental results and analytical models, we conclude that an energetic disorder due to the energy broadening effect of the active layer can be influenced by its adjacent layers remarkably. This may, as a consequence, play a role for the $V_{oc}$ loss. Thus, a careful consideration of the energetic disorder parameter for donor-acceptor organic blends is indeed a valuable criterion in organic BHJ photovoltaic architectures.

**Supplementary Materials**

See supplementary materials for the schematic drawings of the *DOS* (S-Fig.1), photovoltaic *J-V* characteristics (S-Fig.2), schematic diagrams for the molecules and device structure (S-Fig.3), and the illumination intensity dependence of Nyquist plots (S-Fig.4 – 7). S-TAB.1. shows the photovoltaic parameters. S-TAB.2 – 7 show the corresponding fitting parameters for the Nyquist plots.

**ACKNOWLEDGMENTS**

This work was supported by the National Natural Science Foundation of China (Grant No. 61604010, 61634001, U1601651), and the research funding from Beijing Jiaotong University Research Program (Grant No. 2015RC093).

**FIG. 1.**

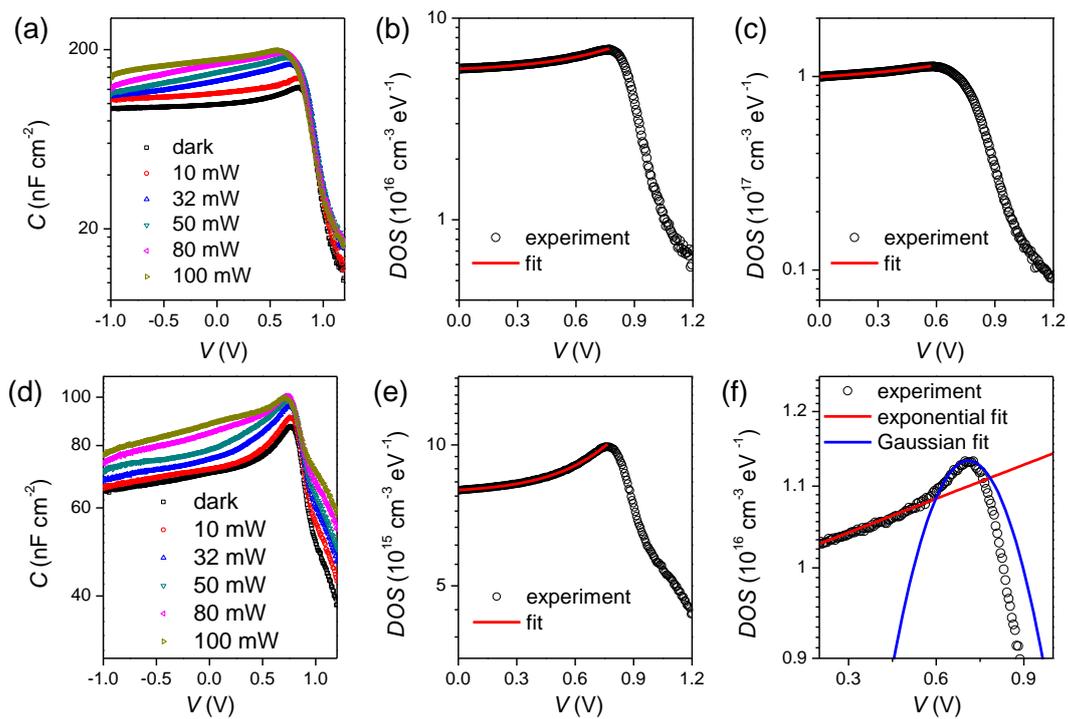

**FIG. 2.**

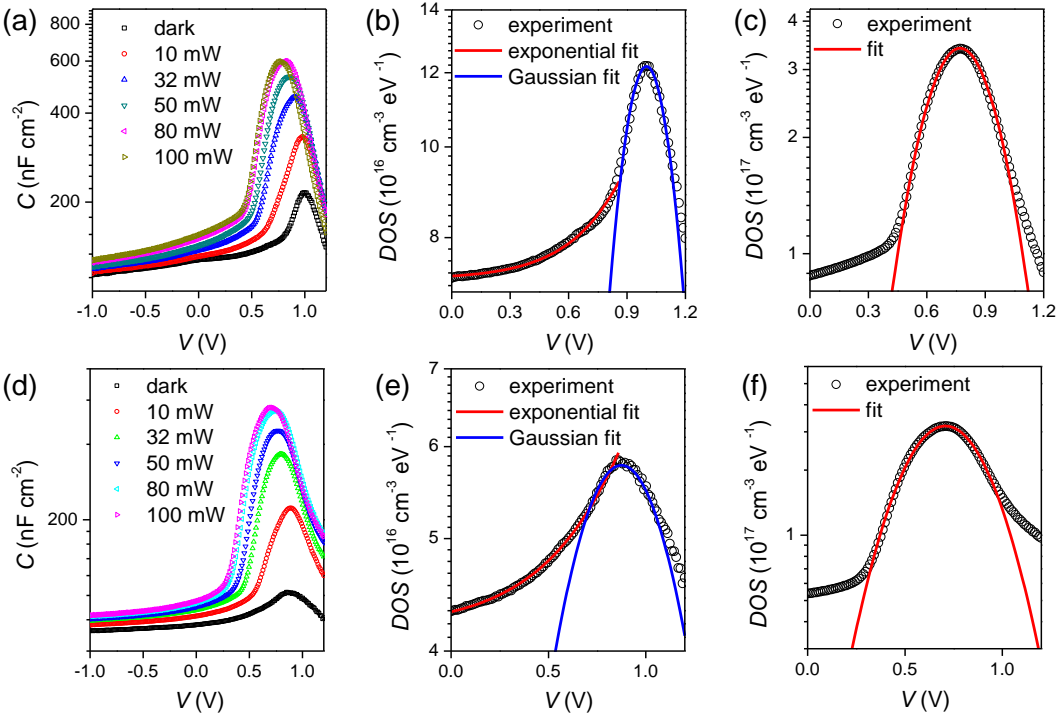

**FIG. 3.**

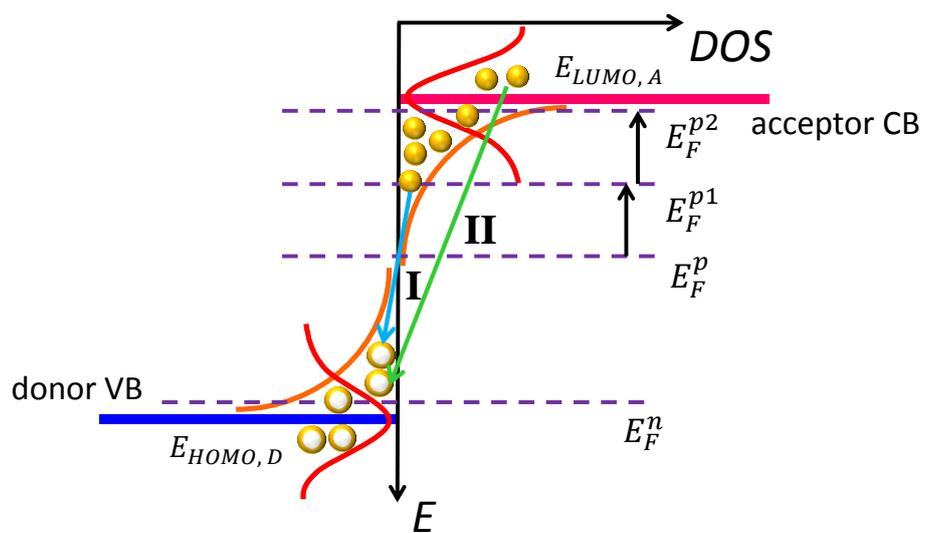

**Figure Captions**

**FIG. 1.** (a) *C-V* measurements at different illumination intensities for an organic BHJ solar cell consists of ITO(glass)/PEDOT:PSS/PIDTDTQx:PC$_{70}$BM/PFN/Al, (b) and (c) are the two selected *C-V* curves measured in dark and under 100 mW/cm$^2$ illumination with the fitting curves (red) respectively. (d) *C-V* measurements at different illumination intensities for an organic BHJ solar cell consists of ITO(glass)/PEDOT:PSS/PIDTDTQx:PC$_{70}$BM/Al, (e) and (f) are the two *C-V* curves measured in dark and under 100 mW/cm$^2$ illumination with the fitting curves (red) respectively.

**FIG. 2.** (a) *C-V* measurements at different illumination intensities for an organic BHJ solar cell consists of ITO/PIDTDTQx:PC$_{70}$BM/PFN/Al, (b) and (c) are the corresponding *C-V* curves measured in dark and under 100 mW/cm$^2$ illumination with the fitting curves (red) respectively. (d) *C-V* measurements at different illumination intensities for organic BHJ solar cells consist of ITO/PIDTDTQx:PC$_{70}$BM/Al，(e) and (f) show the corresponding *C-V* curves measured in dark and 100 mW/cm$^2$ illumination with the fitting curves (red) respectively.

**FIG. 3.** The schematic drawing shows donor valence band (donor VB) and acceptor conduction band (acceptor CB), Gaussian *DOS* (red), exponetional *DOS* (orange), I indicates tail to tail recombination, and II indicates free electron to tail recombination process.

# Supplementary Information

# Unraveling Energetic Disorder in Organic Bulk Heterojunction Photovoltaics by Capacitance-Voltage Spectroscopy

Xixiang Zhu[1], Kai Wang[1]*, Changfeng Han[1], Qin Yang[1], Xiaojuan Sun[1], Haomiao Yu[1], Ming Shao[1], Fujun Zhang[1] and Bin Hu[1, 2]*

3. Key Laboratory of Luminescence and Optical Information, Ministry of Education, Beijing Jiaotong University, Beijing 100044, China
4. Department of Materials Science and Engineering, University of Tennessee, Knoxville, TN 37996, USA

Correspondence should be addressed to:
Kai Wang
E-Mail: kaiwang@bjtu.edu.cn
Bin Hu
E-Mail: bhu@utk.edu

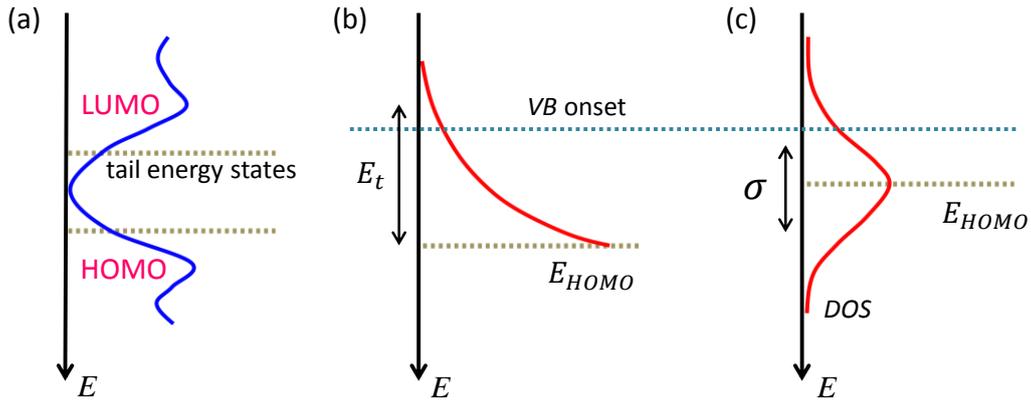

**S-FIG.1.** (a) The schematic drawing of density of states (*DOS*) for a disordered organic semiconductor with tail energy states extending into the bandgap. (b) is an exponential *DOS* and its corresponding width is $E_t$. (c) displays a Gaussian shaped *DOS* and its corresponding width is $\sigma$.

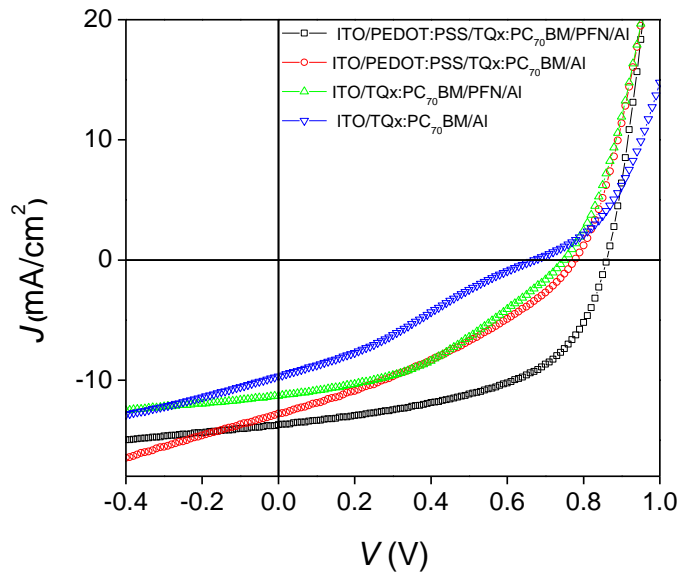

**S-FIG.2.** Photovoltaic *J-V* curves for four different BHJ device configurations, ITO/PEDOT:PSS/PIDTDTQx:PC$_{70}$BM/PFN/Al (black), ITO/PEDOT:PSS/PIDTDTQx:PC$_{70}$BM/Al (red), ITO/PIDTDTQx:PC$_{70}$BM/PFN/Al (green), and ITO/PIDTDTQx:PC$_{70}$BM/Al (blue).

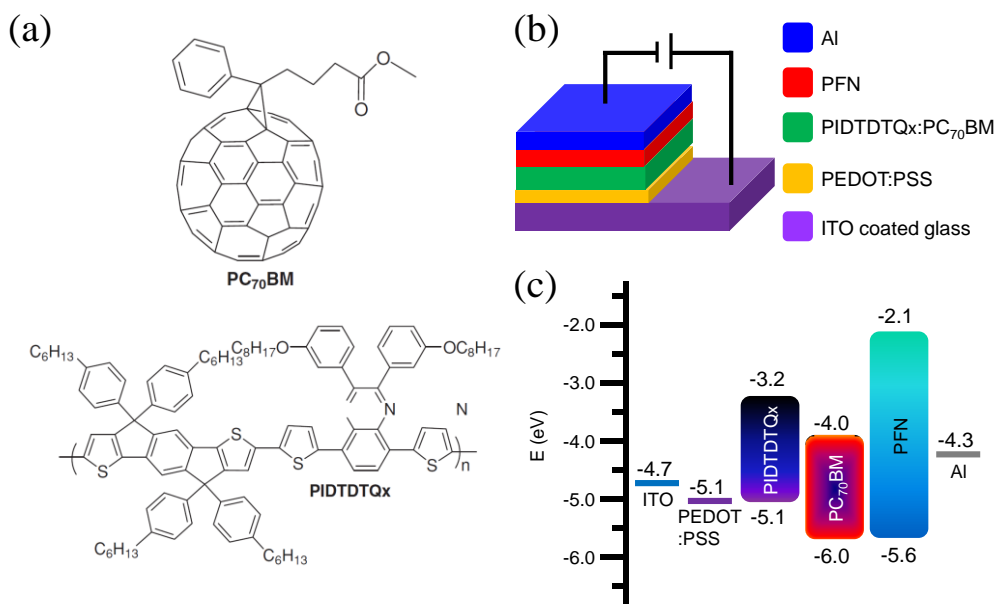

**S-FIG.3.** (a) Schematic diagrams of PC$_{70}$BM and PIDTDTQx molecular structures. (b) Three-dimensional (3-D) view of the device structure. (c) An energy-level diagram for ITO (glass)/PEDOT:PSS/PIDTDTQx:PC$_{70}$BM/PFN/Al.

**Device Fabrication**

Indium tin oxide (ITO) coated glass substrates were cleaned in an ultrasonic bath by standard chemical means. The substrates were dried using pure-nitrogen gas and subsequently were treated by oxygen plasma for 10 min in an enclosed chamber. Then, PEDOT:PSS (Clevios P VP.Al 4083, purchased from H.C. Starck co. Ltd.) films of 40 nm thick were spin-coated onto ITO/glass substrates at 5000 rounds per minute (RPM) for 40 s. The samples were annealed at 150 °C for 15 min in ambient. Afterwards, they were immediately transferred into a pure nitrogen-filled glove-box. PIDTDTQx (i.e.,poly{[4,9-dihydro-4,4,9,9-tetra(4-hexylbenzyl)-s-indaceno[1,2-b:5,6-b´]-dithiophene-2,7-diyl]-alt-[2,3-bis(3-

(octyloxy)phenyl)-2,3-dihydro-quinoxaline-2,2′-diyl], Product No: OT51501, purchased from Organtec Materials Inc.) and PC$_{70}$BM (i.e., [6,6]-phenyl-C-71-butyric acid methyl ester, Product No: OS0633, purchased from 1-Material) with a weight ratio of 1:4 were dissolved in an organic solvent DCB at a concentration of 50 mg/ml, respectively. Blend solutions were placed on a hot plate and stirred using a magnetic stirrer for 12 hours at 70 °C. The previous study has shown the donor PIDTDTQx has high thermal stability with decomposition temperature of 440 °C and its bandgap is approximately equal to 1.88 eV.[1] It also has good solubility in some common organic solvents such as toluene, chloroform, chlorobenzene, and o-dichorobenzene. PIDTDTQx:PC$_{70}$BM weight ratios, indeed, have significant impact on both electron and hole mobilities. With the weight ratio of 1:4, the electron and hole mobilities are about $\mu_e = 4.75 \times 10^{-5}$ cm$^2$V$^{-1}$s$^{-1}$ and $\mu_h = 1.18 \times 10^{-4}$ cm$^2$V$^{-1}$s$^{-1}$ respectively. After this, active layers were spin-coated onto PEDOT:PSS/ITO/glass substrates with a spin-speed of 2500 rpm for 40 s, which produced a film thickness of 110 nm. Prior to the preparation of PFN films, the samples were naturally dried in the nitrogen-filled glove-box for 40 min. The PFN layers of less than 10 nm thick were made with a spin-speed of 3000 rpm for 60 s. Finally, all samples were transferred into an integrated thermal evaporation system; 100 nm Al top contacts were deposited on top of ITO coated glass/PEDOT:PSS/PIDTDTQx:PC$_{70}$BM/PFN samples forming cross-bar structures. The same fabrication method was applied for device structures without PEDOT:PSS or PFN, or both of them. Here, we need to emphasis that both PEDOT:PSS and PFN are the conducting and insulating polymers respectively, and they do not possess any photo-generated-electrical responses. Therefore, electronic transport signals are mainly due to the active layer (i.e., PIDTDTQx:PC$_{70}$BM).

**Electronic Transport Measurements**

Electronic transport measurements were carried out in ambient, current-voltage (*I-V*) characteristics was measured using the two-wire method by a source-meter unit (Keysight B2912A). *C-V* measurements, without and with different illumination intensities through different optical density (OD) filters up to the simulated air mass 1.5 global (AM 1.5 G) condition, were performed by an LCR impedance analyzer (Keysight E4990A) under an alternating electric field of 50 mV (i.e., $V_{peak}$) at the frequency of 10 kHz. For all electronic transport measurements, we adopted that positive (i.e., forward) bias voltages correspond to electrons flow from occupied electronic states of Al into unoccupied electronic states of ITO, and it is vice-versa for negative (i.e., reverse) bias voltages.

| Device Structures | $V_{oc}$ (V) | $J_{sc}$ (mA/cm$^2$) | FF (%) | PCE (%) |
|---|---|---|---|---|
| ITO/PEDOT:PSS/PIDTDTQx:PC$_{70}$BM/PFN/Al | 0.86 | 13.71 | 52.9 | 6.24 |
| ITO/PEDOT:PSS/PIDTDTQx:PC$_{70}$BM/Al | 0.78 | 12.79 | 34.3 | 3.42 |
| ITO/PIDTDTQx:PC$_{70}$BM/PFN/Al | 0.75 | 11.32 | 39.8 | 3.38 |
| ITO/PIDTDTQx:PC$_{70}$BM/Al | 0.67 | 9.69 | 29.1 | 1.89 |

**S-TAB.1.** Photovoltaic parameters, such as $V_{oc}, J_{sc}, FF,$ and PCE, are summarized for the four different device configurations.

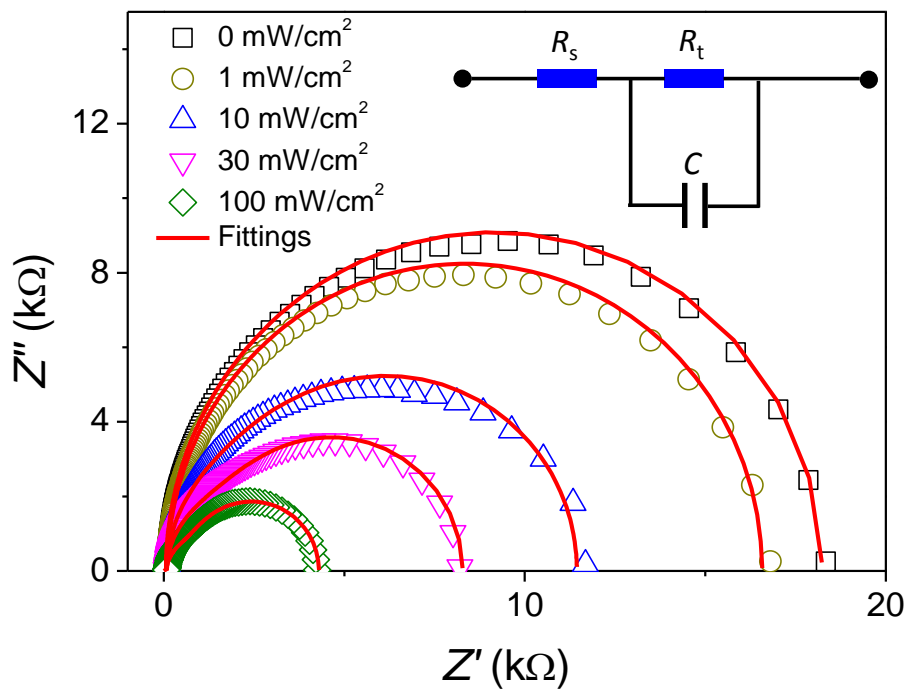

**S-FIG.4.** The illumination intensity dependence of the Nyquist plots for the device consists of ITO/PEDOT:PSS/PIDTDTQx:PC$_{70}$BM/PFN/Al, the fitting curves are denoted by the red solid lines. The inset shows the equivalent electronic circuit.

| Illumination (mW/cm$^2$) | $R_s$ (Ω) | $R_t$ (kΩ) | $C$ (nF) |
|---|---|---|---|
| Dark | 57.55 | 18.1 | 5.52 |
| 1 | 57.6 | 16.7 | 5.45 |
| 10 | 57.9 | 11.6 | 8.88 |
| 30 | 58.1 | 8.2 | 14.02 |
| 100 | 58.35 | 4.2 | 14.93 |

**S-TAB.2.** The fitting parameters are for the impedance spectra of S-FIG.4.

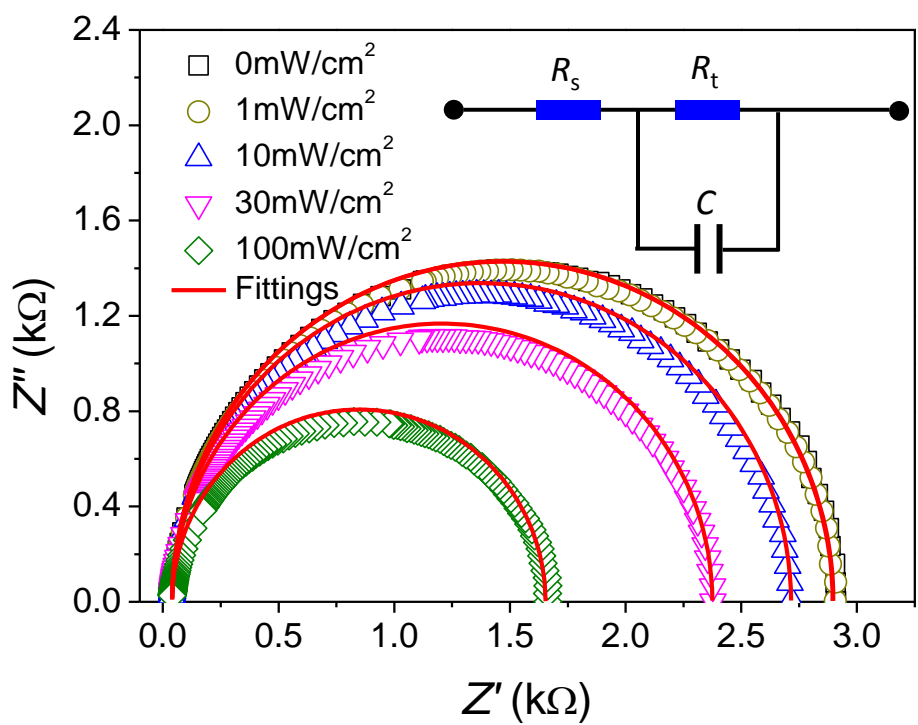

**S-FIG.5.** The illumination intensity dependence of the Nyquist plots for the device consists of ITO/PEDOT:PSS/PIDTDTQx:PC$_{70}$BM/Al, the fitting curves are denoted by the red solid lines. The inset shows the equivalent electronic circuit.

| Illumination (mW/cm$^2$) | $R_s$ (Ω) | $R_t$ (kΩ) | $C$ (nF) |
|---|---|---|---|
| Dark | 39.92 | 2.86 | 3.86 |
| 1 | 39.92 | 2.85 | 4.18 |
| 10 | 39.92 | 2.68 | 6.06 |
| 30 | 39.92 | 2.34 | 7.37 |
| 100 | 39.92 | 1.61 | 8.37 |

**S-TAB.3.** The fitting parameters are for the impedance spectra of S-FIG.5.

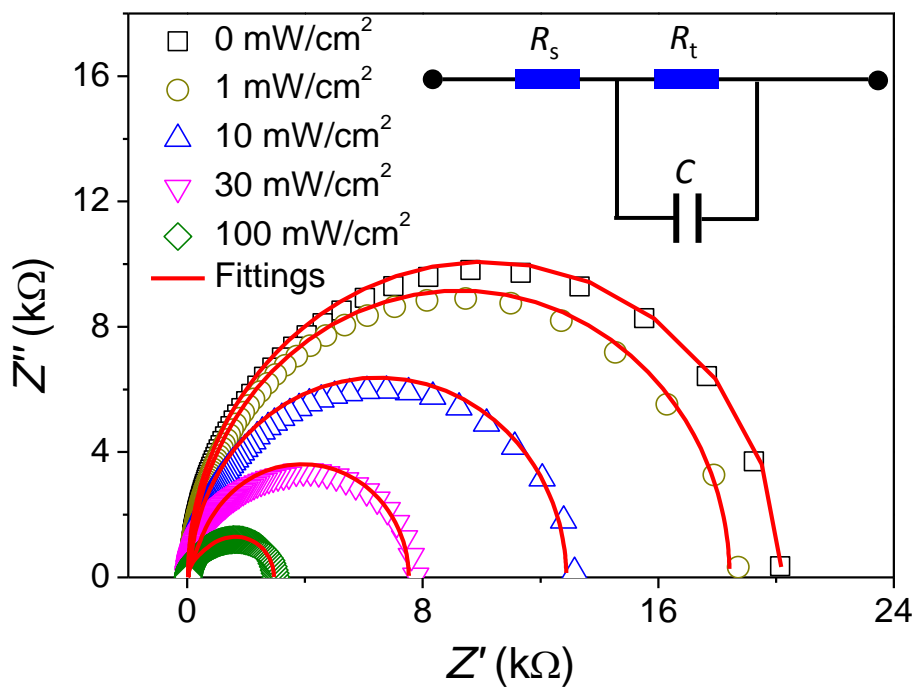

**S-FIG.6.** The illumination intensity dependence of the Nyquist plots for the device consists of ITO/PIDTDTQx:PC$_{70}$BM/PFN/Al, the fitting curves are denoted by the red solid lines. The inset shows the equivalent electronic circuit.

| Illumination (mW/cm$^2$) | $R_s$ (Ω) | $R_t$ (kΩ) | $C$ (nF) |
|---|---|---|---|
| Dark | 26.78 | 20.15 | 6.60 |
| 1 | 26.85 | 18.74 | 7.27 |
| 10 | 26.85 | 13.16 | 8.19 |
| 30 | 27 | 7.66 | 7.32 |
| 100 | 28 | 2.96 | 9.32 |

**S-TAB.4.** The fitting parameters are for the impedance spectra of S-FIG.6.

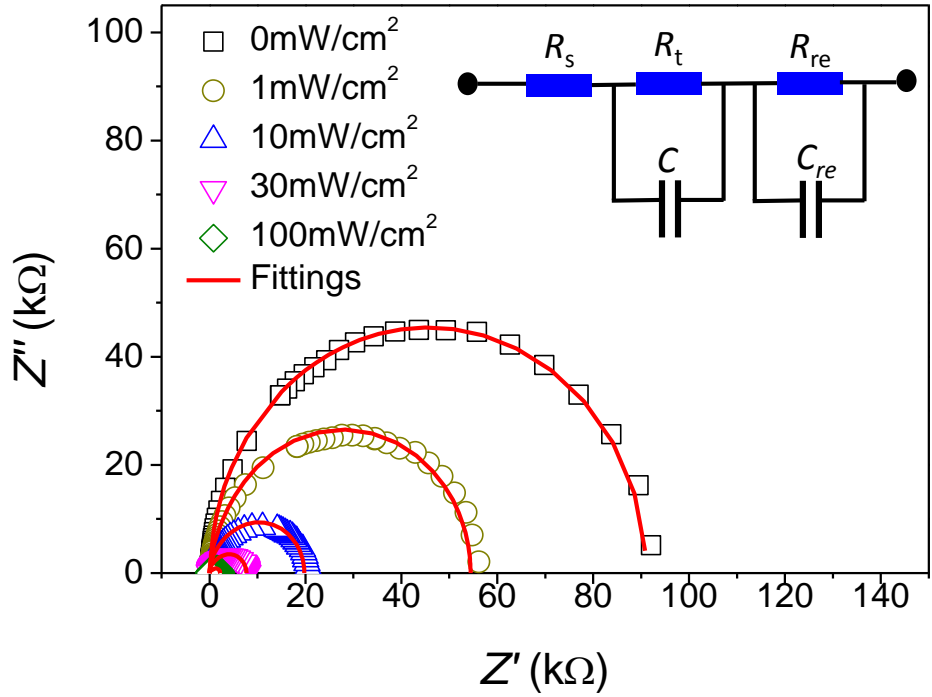

**S-FIG.7.** The illumination intensity dependence of the Nyquist plots for the device consists of ITO/PIDTDTQx:PC$_{70}$BM/Al, the fitting curves are denoted by the red solid lines. The inset shows the equivalent electronic circuit.

| Illumination (mW/cm$^2$) | $R_s$ (Ω) | $R_t$ (kΩ) | $R_{re}$ (kΩ) | $C$ (nF) | $C_{re}$ (nF) |
|---|---|---|---|---|---|
| Dark | 35.71 | 90.7 | 0.17 | 3.96 | 17.71 |
| 1 | 35.82 | 52.5 | 0.22 | 4.12 | 16.06 |
| 10 | 36.06 | 18.9 | 0.39 | 4.72 | 11.81 |
| 30 | 36.47 | 0.71 | 6.85 | 7.61 | 6.45 |
| 100 | 36.88 | 0.58 | 1.57 | 5.14 | 1.32 |

**S-TAB.5.** The parameters are obtained from fitting the impedance spectra of S-FIG.7.